\documentclass[aps,prl,showpacs,twocolumn]{revtex4}
\usepackage{graphicx}
\usepackage{dcolumn}
\usepackage{bm}
\input{epsf}

\begin{document}

\title{Tunneling Study of the Charge-Ordering Gap on the Surface of La$_{0.350}$Pr$_{0.275}$Ca$_{0.375}$MnO$_3$ Thin Films}

\author{Udai Raj Singh, S. Chaudhuri, Shyam K. Choudhary, R. C. Budhani, and Anjan K. Gupta}
\affiliation{Department of Physics, Indian Institute of Technology
Kanpur, Kanpur 208016, India.}
\date{\today}

\begin{abstract}
Variable temperature scanning tunneling microscopy/spectroscopy studies on (110) oriented epitaxial thin
films of La$_{0.350}$Pr$_{0.275}$Ca$_{0.375}$MnO$_3$ are reported in
the temperature range of 77 to 340 K. The films, grown on lattice
matched NdGaO$_3$ substrates, show a hysteretic metal-insulator
transition in resistivity at 170 K. The topographic STM images show
step-terrace morphology while the conductance images display a
nearly homogeneous surface. The normalized conductance spectra at
low temperatures (T$<$150 K) show an energy gap of 0.5 eV while for
T$\geq$180 K a gap of 0.16 eV is found from the activated behavior
of the zero bias conductance. The presence of energy gap and the absence of
phase separation on the surface over more than
2 $\mu$m$\times$2 $\mu$m area contradicts the metallic behavior seen
in resistivity measurements at low temperatures. We discuss the
measured energy gap in terms of the stabilization of the insulating
CO phase at the film surface.
\end{abstract}


\maketitle
The research over past decade on colossal magneto-resistive (CMR) manganites
\cite{book-rao-raveau,salamon-jaime,von-helmolt} has led to a number
of very interesting phenomena. These are now understood to be
systems with strongly coupled charge, spin, lattice, and orbital
degrees of freedom with a number of possible phases of nearly equal
free energies \cite{dagotto-rev,salamon-jaime,ahn}. These phases are
thought to coexist in some manganite compositions and the relative fractions
of phase are sensitive to external perturbations such
as magnetic field, electric field, pressure, and strain. In fact,
a sudden termination of the periodic potential at the surface is a
perturbation that seems to make the surface quite different from the
bulk \cite{freeland-plummer}. The bulk of some of the small
band-width manganites \cite{hwang-batlogg} undergoes an electronic
phase-separation \cite{uehara-cheong,kim-cheong} into metallic and
insulating regions, which is thought \cite{dagotto-rev,moreo,ahn} to
be responsible for the CMR effect. Various research groups have
observed multiple phases in some of the manganites, in particular,
in (La$_{1-x}$Pr$_{x}$)$_{0.625}$Ca$_{0.375}$MnO$_3$ series, using
different probes \cite{uehara-cheong,zhang-lozanne,lee} including
Scanning Tunneling Microscopy / Spectroscopy (STM/S)
\cite{fath,renner-cheong,becker,moshnyaga}. However, the detailed
electronic nature of the individual phases is far from understood
\cite{salamon-jaime}.

In this work, we have probed the surface of a relatively narrow
band-width  hole-doped manganite
La$_{0.350}$Pr$_{0.275}$Ca$_{0.375}$MnO$_3$ (LPCMO), using the STM/S
in the temperature range of 77K to 340K.  The ground state of the
two end members of this compound La$_{0.675}$Ca$_{0.375}$MnO$_3$ and
Pr$_{0.675}$Ca$_{0.375}$MnO$_3$ are ferromagnetic metal and
antiferromagnetic charge-ordered (CO) insulator, respectively.
Intermediate compositions show a CO transition around 200 K with
coexisting CO insulating and charge-disordered conducting phases
\cite{uehara-cheong}. Our main objective in this study has been to
probe various phases in LPCMO films by local tunneling spectroscopy
to understand their nature and contribution to CMR. However, we find
that the low energy part of the spectra evolves quite homogeneously
with temperature with an energy gap that changes from 0.16 eV to 0.5
eV on cooling. Whilst the gap varies slightly on a nanometer length scale,
it exists even below the insulator-metal transition temperature ($T_{MI}$) measured
by resistivity. This is the first temperature dependent STM/S study on this narrow bandwidth
manganite, and our major finding is the persistence of a nearly
homogeneous CO gap which we interpret in terms of the stabilization
of the CO phase on the surface even though the bulk of the film
shows a phase separated behavior.

Epitaxial films of thickness $\approx$ 200 nm of LPCMO were grown at
a deposition rate of $\sim$ 0.3 nm/s on (110) surface of NdGaO$_{3}$ (NGO) single
crystals by using pulsed laser deposition (PLD). The (110) direction refers to the orthorhombic
unit cell of NGO with  d$_{110}$ = 0.38595 nm. A KrF excimer laser
\begin{figure}
\epsfxsize = 1.5 in \epsfbox{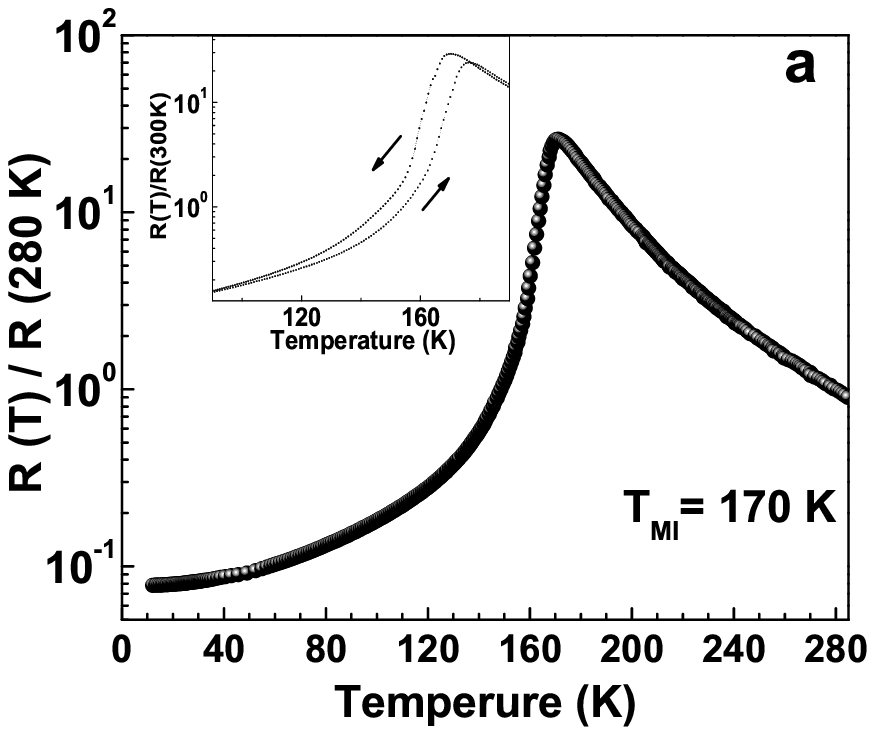}
\epsfxsize = 1.4 in \epsfbox{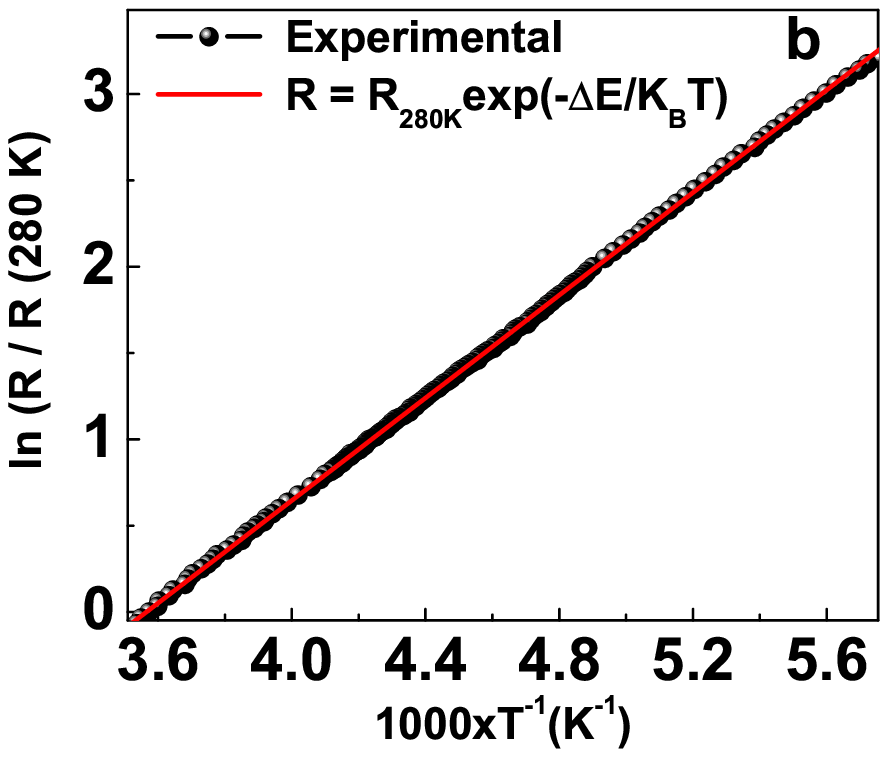}
\caption{\label{fig:res} a.
Normalized resistance as a function of temperature showing the
metal-insulator transition in LPCMO film at 170 K. The inset in fig.1(a) showing the hysteric metal-insulator transition.
The plot of ln(R/R$_{280 K}$) with 1000/T (for 280 K$>$T$>$170 K) is
shown in the fig1.(b) with a linear fit giving an activation energy of
0.12 eV.}
\end{figure}
operated at 10 Hz with an areal energy density of 2 Jcm$^{-2}$/pulse
on the surface of a stoichiometric sintered target of LPCMO. For
these films we find a step-terrace morphology for relatively slow
growth conditions while the faster growth rates lead to a granular
surface. In this paper, we focus on the STM/S study of LPCMO thin films with step-terrace surface morphology. Films with granular morphology have an associated electronic inhomogeneity and they may not reflect the film's intrinsic electronic behavior. For resistivity measurements, silver contact pads were deposited through a shadow mask in a four probe configuration. The STM/S
measurements were performed using a homemade variable temperature
STM with fresh cut Pt$_{0.8}$Ir$_{0.2}$ wire tip. The STM head is
based on a design published elsewhere \cite{anjan-stm} while the
electronics and software are commercial. The STM measurements
reported here have been carried out in liquid nitrogen in 77 - 340 K
temperature range in a cryogenic vacuum. After transferring the
sample to the STM in air, the cryostat is pumped and flushed with
He-gas a few times. The sample transfer is carried out quite fast so that
the sample is exposed to air for less than 30 min. Finally the
cryostat is pumped to a high vacuum (1$\times$10$^{-4}$ mbar) prior to cooling in liquid nitrogen. To ensure a
clean sample surface we heat the whole STM to ~340K for a few hours keeping the surroundings at liquid nitrogen temperatures.

The standard ac-modulation technique was used for STS and
conductance spectra with a modulation amplitude between 30 and 50
mV. The dc-component of current, $I(V)$, is fed to the z-feedback
for controlling and measuring the tip-sample separation while a
lock-in amplifier measured the ac-component giving the conductance,
dI/dV, at bias V. Simultaneous conductance and topographic images in
constant-current mode at a bias (V=$V_0$) are obtained by measuring
the lock-in output and the z-feedback voltage, respectively, over
the surface. In this mode,  the feedback, by varying tip-sample
separation, ensures that the integrated conductance up to $V_0$ is
the same at all points of the image. As a result, in the dark
regions of a conductance image the current gets less contribution
from the conductance at larger bias and more contribution from the
small bias. Thus, in a conductance image, the darker regions
represent more metallic (or less gapped) areas \cite{renner-book-ch}
and the topographic and conductance images are anti-correlated for
flat surfaces.

The four probe resistance as a function of temperature for an LPCMO
film (Fig.\ref{fig:res}a) shows an insulator-metal transition at
T$_{MI}$ = 170 K with more than two orders of magnitude change in
resistivity, which is typical of the narrow bandwidth CMR manganites
\cite{wu-greene-millis}. The cooling and heating curves differ
slightly near the transition indicating the presence of more than
one phase in the bulk of this film as shown in the inset of Fig.\ref{fig:res}a. Resistivity in LPCMO single
crystals and polycrystalline samples shows a small feature at
$\sim$200 K due to opening of a CO gap \cite{uehara-cheong,lee},
which is absent in thin films \cite{Tara Dhakal, W. Prellier}. This
could be due to the small substrate induced strain effect making the
CO transition not so sharp with respect to temperature. The formation of charge ordered state below 210 K in the same composition
thin films on NGO substrate has been observed by electron microscopy  and diffraction studies by Chaudhuri, et. al.\cite{Chaudhuri}. In Fig.\ref{fig:res}b, we plot ln($R/R_{280K}$) Vs 1/T for 170
K$<$T$<$285 K, which establishes the activated behavior of
resistivity with an activation energy ($\Delta$) of 0.12 eV. This is
in close agreement with the average optical gap (2$\Delta$) of about
0.2 eV in this temperature range observed by Lee et.al. \cite{lee}
in LPCMO single crystals of slightly different composition.

\begin{figure}
\epsfxsize = 3.3 in \epsfbox{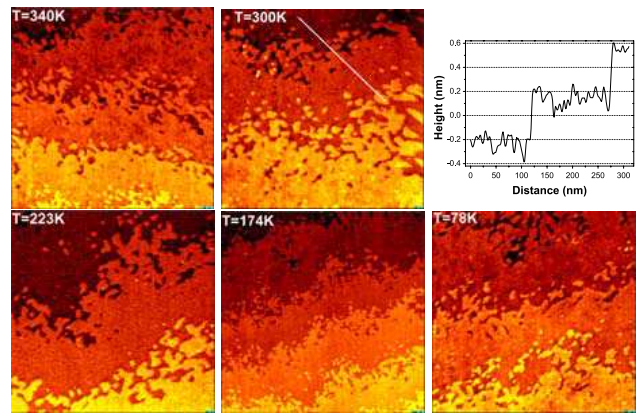}
\caption{\label{fig:topoLTRT}Topographic images of LPCMO film at
temperatures of 340 K (Scan size = 800$\times$800 nm$^2$, Bias = 1 V / Current = 0.1 nA), 300 K (502$\times$502 nm$^2$, 1.1 V / 0.1 nA), 223 K (504$\times$504 $nm^2$, 1.1 V / 0.1 nA), 174 K (1.5$\times$1.5 $\mu$m$^2$, 1.4 V /
0.2 nA), and 78 K (700$\times$700 nm$^2$, 2.6 V / 0.3 nA) The graph in the top
right shows a line cut along the line marked in 300 K image with
steps of height 0.4 ($\pm$0.05) nm.}
\end{figure}

The STM topographic images of the LPCMO film (in
Fig.\ref{fig:topoLTRT}) between 340 and 78 K temperature show
step-terrace morphology with a terrace width of about 200 nm. The
rms roughness of each terrace is found to be less than 0.1 nm. From
the conductance images, as discussed later, some of this roughness
can be attributed to the electronic inhomogeneities. The terrace
boundaries are quite irregular together with a few isolated islands
of lateral size up to ten nanometers. From the line-cut shown in
Fig.\ref{fig:topoLTRT}c, the height of the terrace is found to be
0.4 ($\pm$0.05) nm in agreement with the step height of 0.38 nm for
(110) planes of LPCMO. Further, a tiny (0.2\%) epitaxial mismatch
between the NGO substrate and the LPCMO films for such large
thickness would not leave significant strain near the surface. The
presence of such smooth terraces allows local tunneling experiments
in the regions that are unaffected by the presence of steps or grain
boundaries and their associated local disorders and
non-uniformities.

\begin{figure}
\epsfxsize = 1.22 in \epsfbox{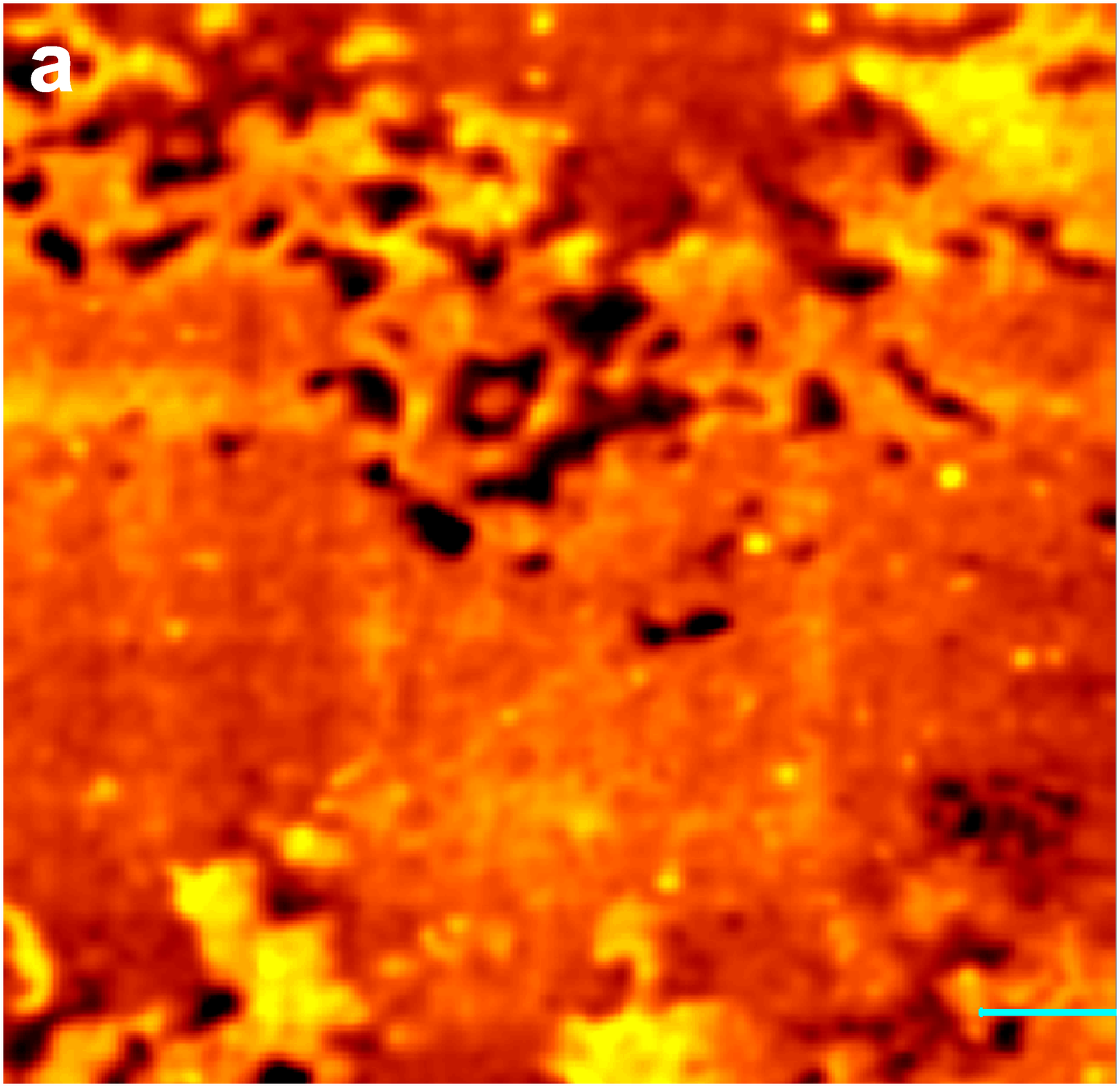}
\epsfxsize = 1.2 in \epsfbox{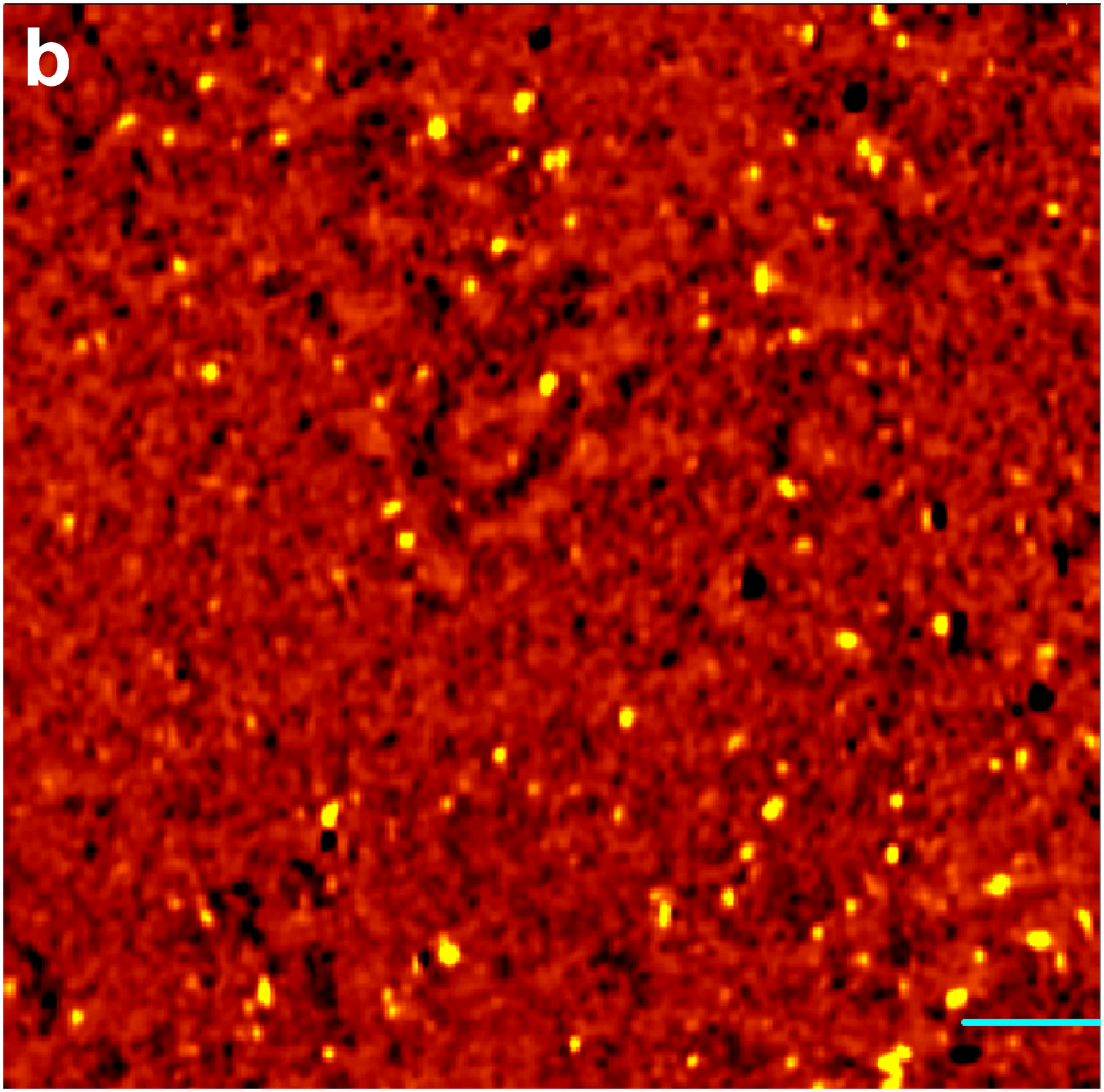}
\epsfxsize = 1 in \epsfbox{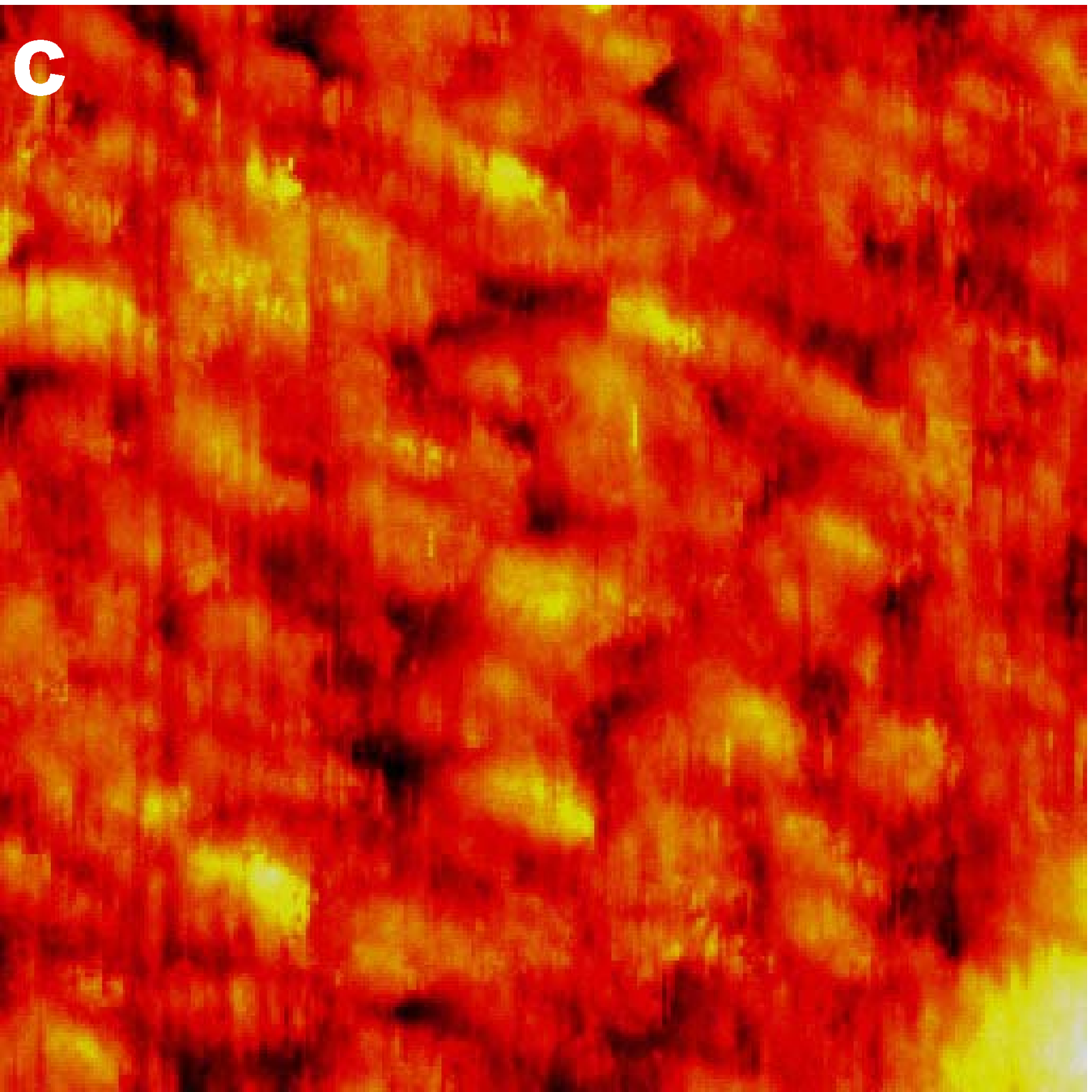}
\epsfxsize = 1 in \epsfbox{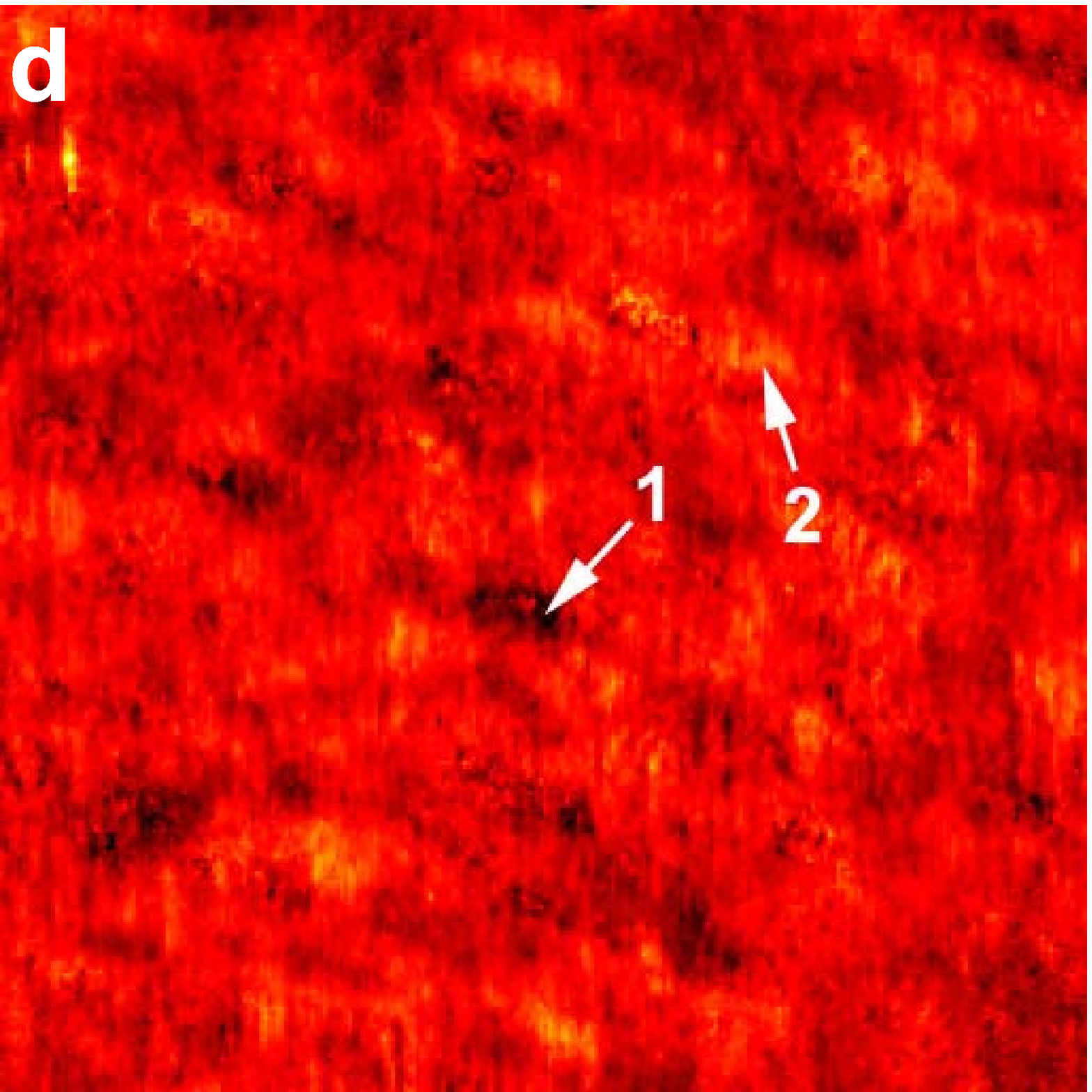}
\epsfxsize = 1.2 in \epsfbox{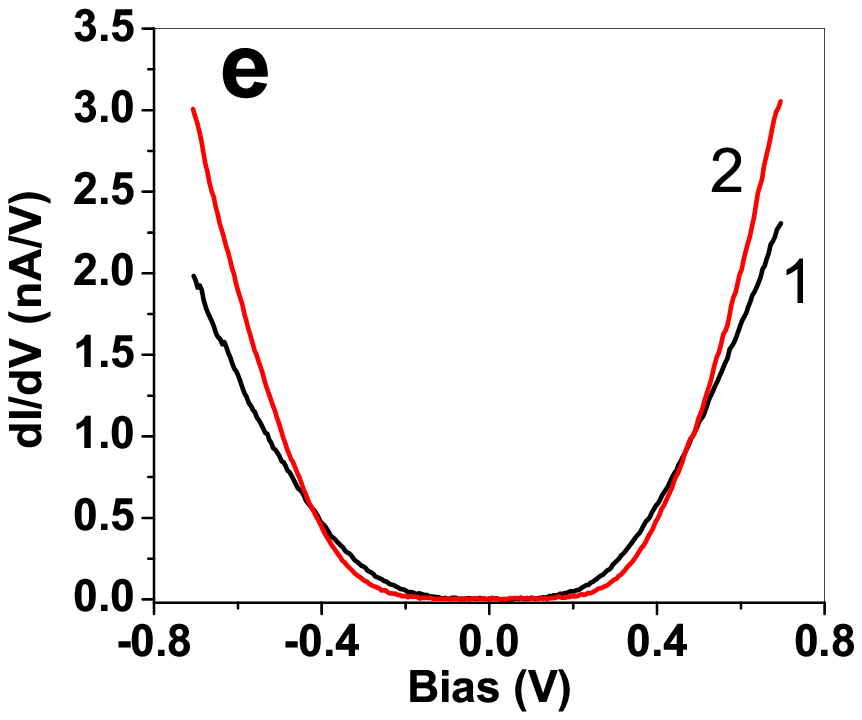}
\caption{\label{fig:STS-images} Topographic (a and c) and
corresponding conductance (b and d) images at T = 140 K at bias 1.0 V and tunnel current value
0.1 nA. The size of images a and b is 416$\times$416 nm$^2$ and that of
c and d is 46$\times$46 nm$^2$. fig.(e) shows characteristic spectra at
two representative points (dark and bright) of the image in d.}
\end{figure}

Fig.\ref{fig:STS-images} shows a set of simultaneously taken
topographic and conductance images at 140 K. As discussed earlier,
the bright regions of a conductance image represent more gapped
areas. The conductance images have some very bright spots of about 5
nm size with a relatively low coverage (~20 spots/$\mu m^2$). The
image in Fig.\ref{fig:STS-images}b has a number of these bright
spots. These spots are seen at all temperatures and they have a
large energy gap (1 to 1.5 eV) as found from the local spectra. It is
possible that these are either second phase materials belonging to
the parent compounds, which are all antiferromagnetic insulators, or
oxides of the constituent metals. Since their fraction is very small
it is difficult to detect them in X-ray diffraction. Such defects
are also not seen in high resolution TEM images of similar samples,
which makes us believe that they are segregated at the surface.

Other than these large-gap-spots, a weak contrast in the conductance
images (Fig.\ref{fig:STS-images}d) is also seen at nanometer length scale, which is somewhat
anti-correlated with the contrast of the simultaneous topographic
image. The image in Fig.\ref{fig:STS-images}d is a smaller area image of region
without any bright spots as discussed above. The cross-correlation coefficient between the topographic and
conductance images on terraces is found to be $\approx$ -0.2 showing
that the density of states (DOS) inhomogeneity is partly responsible
for the topographic contrast. The detailed tunneling spectra, with
two representative ones in Fig.\ref{fig:STS-images}e, show that this
inhomogeneity actually corresponds to a small variation in the
CO-gap value. Similar inhomogeneity is seen at all temperatures from
77 to 340 K. We believe that this electronic contrast represents
some frozen chemical inhomogeneity in this non-stoichiometric
compound.

\begin{figure}
\epsfxsize = 1.71 in \epsfbox{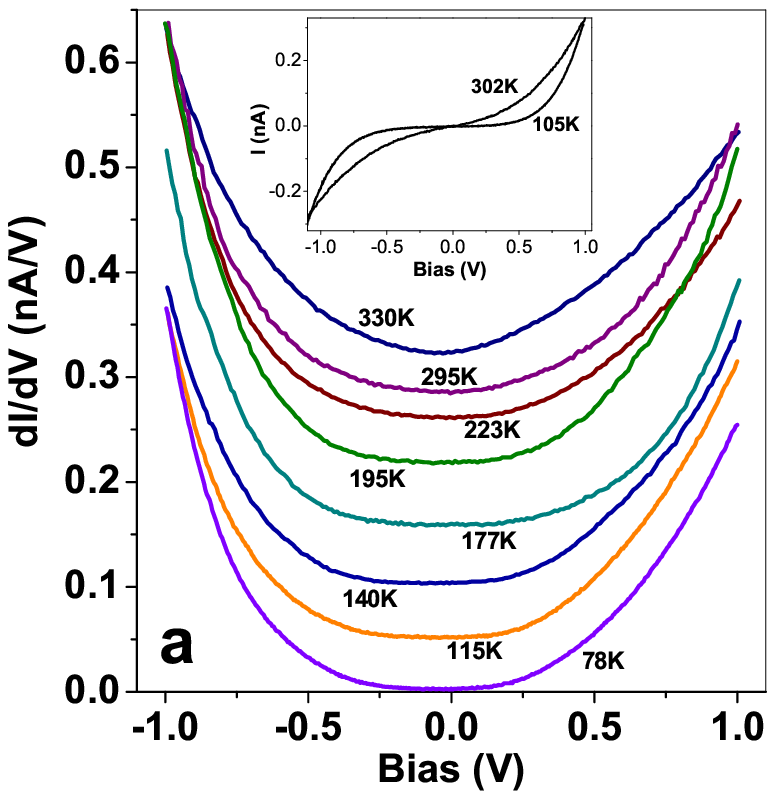} \epsfxsize = 1.61 in
\epsfbox{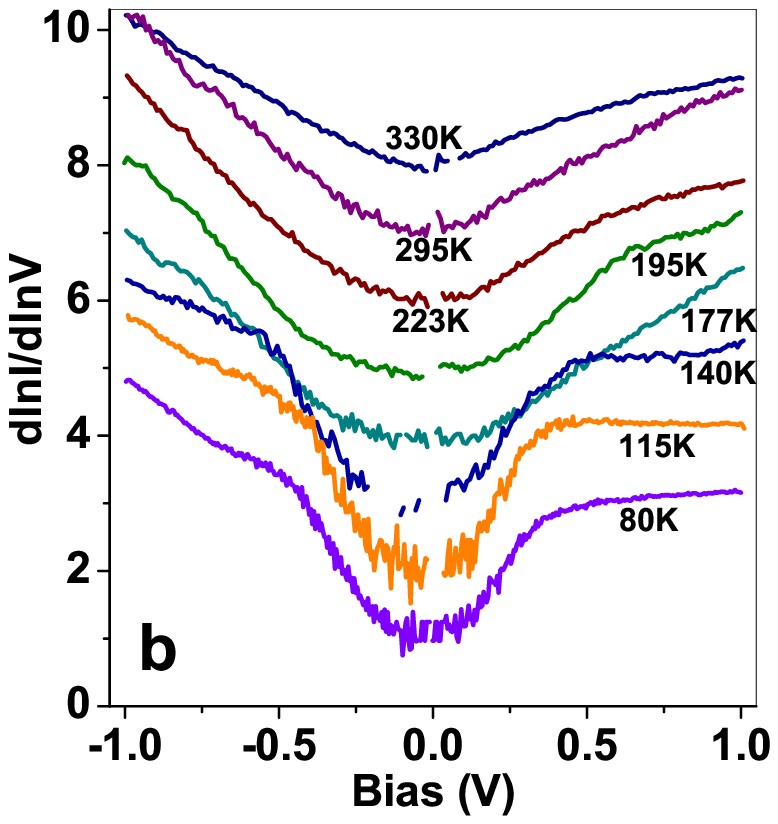} \caption{\label{fig:spec-t-dep} Temperature
dependent dI/dV-V (left) and dlnI/dlnV-V (right) spectra with
junction bias 1.0V and current 0.1nA. I-V spectra at two different
temperature 302 K and 105 K are shown in the inset of the left plot.
The consecutive spectra have been shifted upward for clarity.}
\end{figure}

A tunneling spectrum, other than the DOS, is also affected by finite
temperatures and the voltage dependence of the tunneling matrix
element \cite{wolf,stroscio-feenstra}. The temperature smears out
the spectral features of width less than a few $k_BT$. The voltage
dependence of the matrix element is more difficult to handle for
large bias spectra as one does not know much about the details of
the tunnel barrier. In general, one has to be careful in comparing
the absolute dI/dV values at a given bias for two spectra that have
different behavior with V as it is difficult to ensure the same
tunneling matrix element for the two spectra. For such comparisons we
have chosen the spectra with same junction resistances at a fixed
bias. To extract the DOS from a dI/dV spectra a widely used method
\cite{stroscio-feenstra} is to plot (dI/dV)/(I/V) (or
d(lnI)/d(lnV)), which normalizes away the voltage dependence of the
matrix element from the spectra. Practically, d(lnI)/d(lnV) sharpens
the spectral features that are not smeared away by temperature.

Temperature dependence of the tunneling spectra is shown in
Fig.\ref{fig:spec-t-dep}a with each spectrum taken at a tunnel
current set-point of 0.1 nA at 1 V bias. The conductance near zero
bias at low temperatures is nearly flat with a very small magnitude
indicating the presence of a gap while that at higher temperatures
has a noticeable curvature with significant conductance value. This
indicates opening of an energy gap with cooling. The low temperature
gap is more clearly visible in dlnI/dlnV-V plots in
Fig.\ref{fig:spec-t-dep}b. We also show two representative I-V
spectra in the inset of Fig.\ref{fig:spec-t-dep}a. The gap value
(2$\Delta$) at low temperatures is found to be 0.5($\pm$0.1) eV from
the separation between the center of the two gap edges in dlnI/dlnV
spectra (Fig.\ref{fig:spec-t-dep}b). This is in reasonable agreement
with the low temperature gap of 0.38 eV found in optical
measurements of Lee et.al. \cite{lee}. As mentioned earlier, this
gap value has small ($\approx$ 0.1 eV) variations over a few
nanometer length scale but the spectra are qualitatively
homogeneous. To look for the phase separation, we have studied the
spectra over reasonably large areas ($>$2 $\mu$m$\times$2 $\mu$m) with
step-terrace morphology without any signature of non-gapped spectra
at low temperatures. We have also thermally cycled this sample
several times but this behavior remains unchanged.
\begin{figure}
\epsfxsize = 2.2 in \epsfbox{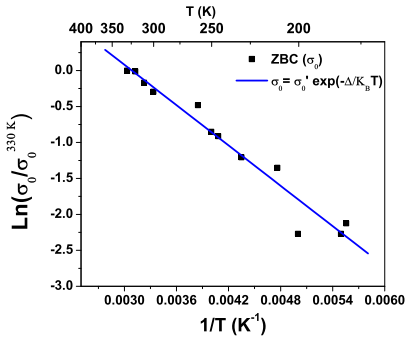} \caption{\label{fig:ZBC}
ln($\sigma_0$) vs 1/T showing an activated behavior of the ZBC
with an energy gap ($\Delta$) of 0.08 eV. The temperature scale is shown on top of X-axis. }
\end{figure}
At higher temperatures, the zero bias conductance (ZBC), $\sigma_0$,
is found to show an activated behavior indicative of the presence of
a smaller gap that is thermally smeared out. Ln($\sigma_0$) above
180 K is plotted in Fig.\ref{fig:ZBC} with respect to 1/T. The
linear fit shown in the same figure gives an activation gap
($\Delta$) of 0.08 eV.

Phase separation into a charge ordered insulating and a charge
disordered conducting phase on sub-micron scale is well established
in LPCMO bulk as well as thin films. This is also apparent in thin
films studied here from the hysteretic behavior of resistivity. The
charge-order exists in LPCMO both below and above $T_{MI}$
\cite{kiryukhin-jeong} with a change in the CO fraction with
temperature and magnetic field \cite{uehara-cheong}. The insulating
behavior above $T_{MI}$ of the LPCMO films studied here can be
attributed to the presence of the CO phase. We found an energy gap
($\Delta$) above $T_{MI}$ of magnitude 0.12 eV from the resistivity
measurements while the gap found from the ZBC of the tunneling
spectra is 0.08 eV. These two are consistent with each other given
that one probes the surface while the other probes the bulk of the
film. An energy gap (2$\Delta$) of 0.5 eV below $T_{MI}$ is found
from dlnI/dlnV spectra. Both these gaps agree reasonably well with
those observed by optical spectroscopy \cite{lee}.

The tunneling measurements above $T_{MI}$ agree with the bulk
resistivity but the two show quite opposite behavior at low
temperatures. It is impossible to get such conducting behavior in
bulk at low temperatures with such a pronounced gap in the
electronic DOS in this three dimensional compound. This result is
extremely puzzling and the only reconciliation we can think of is
that the tunneling somehow is not probing the bulk properties of the
film. Surface contamination is the first thing to blame and we
cannot rule this out completely as we do expose our samples to air
but for a very short time while transferring them from the growth
chamber to the STM. However, excellent terraces over large areas and
a good correlation of the temperature dependent tunneling data with
charge-ordering makes us strongly believe that we are probing
intrinsic properties of the surface of the films, which has a
charge-ordered phase stabilized.

The X-ray spectroscopy and spin polarized tunneling studies on manganites
\cite{freeland-plummer} also show that the surface layer is
insulating with significant suppression in ferromagnetism. The surface is also found to be different from the bulk from the NMR studies on polycrystalline manganites \cite{bibes}. In LPCMO,
this effect could be quite strong as it is a three dimensional
compound. So it is reasonable to assert  that a sudden termination
of the lattice stabilizes a charge-ordered insulating phase on the
surface. Surface reconstruction or a different chemical composition
of the film, particularly oxygen stoichiometry, near the surface are
the two possibilities that we can speculate. Nevertheless, this
seems to be an intrinsic feature of manganites and it will play a
significant role in polycrystalline samples \cite{grain-tunneling,
sudheendra-rao} and devices utilizing the spin-polarization
\cite{planar-junction} of the surface electrons.

In conclusion, our STM/S studies on epitaxial
La$_{0.350}$Pr$_{0.275}$Ca$_{0.375}$MnO$_3$ films on NdGaO$_3$
substrate show a temperature dependent energy gap. Conductance
images show some inhomogeneity on nanometer length scale; however,
no sign of any metallic phase was found in the tunneling spectra for
T$<$T$_{MI}$. The low temperature gap seen here is {\it
inconsistent} with the resistivity measurements which shows a
metallic behavior below T$_{MI}$. From the correlation of the energy
gap with the CO phase we conclude that a charge-ordered insulating
phase remains stabilized near the surface, which masks the
signatures of the phase separation in the bulk of the film.

\section{Acknowledgements}
URS acknowledges financial support from CSIR, AKG acknowledges a
research grant from DST of the Govt. of India. S. Rajauria's
contribution to the initial work on the STM set-up is acknowledged.

\end{document}